\documentclass{acta_hungarica_preprint}
\usepackage{bm}

\usepackage{calrsfs}
\DeclareMathAlphabet{\pazocal}{OMS}{zplm}{m}{n}

\usepackage{color}

\usepackage{caption}
\usepackage{subcaption}

\usepackage{xfakebold}
\usepackage{pgfplots}
\pgfplotsset{compat=newest}
\usepackage{tikz}
\usepackage{tikzscale}
\usepackage{placeins}

\usepackage{array}
\usepackage{booktabs}
\usepackage{multirow}

\newcommand{\vek}[1]{{\mathbf #1}}

\newcommand{\sv}[1]{\setBold[0.3] #1 \unsetBold}

\newcommand{\play}[1]{^{(#1)}}

\newcommand{\APlay}{\pazocal{P}}
\newcommand{\ert}[1]{\eqref{#1}}

\newcommand\figHight{5.6}

\begin{document}

%start page set by editors, leave blank
\setcounter{page}{1}
%title of the paper Model-Based Shared Control Using LQ-Differential Games
\title{Toward Adaptive Cooperation: Model-Based \\ Shared Control Using LQ-Differential Games}
\shorttitle{Adaptive Model-Based Shared Control}
\author[]{Balint Varga}
\affil[]{Institute of Control Systems, Karlsruhe Institute for Technology, 76131 Karlsruhe, Germany, balint.varga2@kit.edu}

\shortauthor{Balint Varga}

\maketitle

\begin{abstract}
This paper introduces a novel model-based adaptive shared control to allow for the identification and design challenge for shared-control systems, in which humans and automation share control tasks. The main challenge is the adaptive behavior of the human in such shared control interactions. Consequently, merely identifying human behavior without considering automation is insufficient and often leads to inadequate automation design. Therefore, this paper proposes a novel solution involving online identification of the human and the adaptation of shared control using Linear-Quadratic differential games. The effectiveness of the proposed online adaptation is analyzed in simulations and compared with a non-adaptive shared control from the state of the art. Finally, the proposed approach is tested through human-in-the-loop experiments, highlighting its suitability for real-time applications.
\end{abstract} 

\vspace*{-8mm}

\section{Introduction}
It has long been known that humans are not always able to perform technical tasks perfectly \cite{2016_SharedControlSharp_flemisch}. In consideration of current technological developments, the obvious solution to this problem seems to be the replacement of the human operator with autonomous systems. The human would only have to monitor these autonomous systems. However, experiences across various domains reveal inherent problems with this approach \cite{1983_IroniesAutomation_bainbridge}. For instance, in automated aircraft cockpits, the "out-of-the-loop" problem is well-documented already in the 1980s, where fully automated systems can lead to inadequate human monitoring \cite{1980_FlightdeckAutomationPromises_wiener}. Recognizing these challenges, cooperative control systems emerge as a promising alternative. In such systems, humans interact and cooperate with automation to perform tasks jointly, necessitating active research in modeling and analyzing these cooperations \cite{2016_SharedControlSharp_flemisch}. 

A broad field is the cooperative shared control for the guidance of passenger cars \cite{9369899, 2023_Na_Cole}. In the vehicle, the steering wheel serves as an interface for lateral guidance, and the pedals serve as interfaces for longitudinal guidance. An analogous problem to cooperative shared control in driver assistance exists in other areas where humans and automation perform tasks cooperatively. These areas include cooperative teleoperation systems \cite{2022_SharedControlRobot_gottardi,2022_IntelligentInteractionFramework_du}, where the operator is supported by assistance systems, cooperative load carrying \cite{2022_ReviewHumanMachine_yanga}, deep-sea dredging \cite{2013_HapticSupportBimanual_kuiper}, aircraft control \cite{2017_NeuromuscularSystemBasedTuningHaptic_smisek}, or mobile manipulators \cite{2023_LimitedInformationShared_varga}. The primary challenge in such cooperative schemes arises when either the human fails to comprehend the automation or vice versa \cite{2012_HapticSharedControl_abbink}. To overcome this challenge, models for human actions and motion primitives are developed to allow the automation to adapt to the human. However, the existing approaches in the literature hinder model-based control design, which is necessary for safety and stability investigations. 

Another approach to modeling human behavior and motion is through optimal control theory \cite{2002_OptimalFeedbackControl_todorov}. This approach assumes that humans optimize an internal cost function, and this optimization leads to a control law explaining human motion. The benefits and results of this modeling approach are discussed in numerous works in the literature, for example, in \cite{1969_HumanOptimalController_baron} and \cite{2002_HumanMotionPlanning_loa}. The inverse approach, which involves identifying such a cost function of the human, is also addressed in the literature, as presented in \cite{2010_HumanHumanoidLocomotion_mombaur} or \cite{2015_SolutionsInverseLQR_priess}.

In \cite{2009_NashEquilibriaMultiAgent_braun}, it is shown that the haptic interactions between humans lead to a Nash equilibrium, indicating the use of game theory to model shared control interactions. Based on this idea, in \cite{2014_SteeringDriverAssistance_flad}, a systematic design of shared control is presented that can optimally support the human operator. For systematic design, an identification of the human cost function is necessary. However, this cost function changes in a shared control configuration, such that the designed controller is no longer optimal. In \cite{2019_SolutionSetsInverse_inga}, the identification of the players' cost function based on an LQ-differential game formulation is proposed. That method was successfully applied in robotic applications to identify the human motion \cite{2022_inga_application,2023_IdentificationHumanControl_,Preferencemodelling_Palo}. 
Current model-based robotics research works overlook a crucial aspect: The human adapts in shared control interaction, which necessitates an online adjustment to the automation. Such an online adaptation\footnote{In this paper, the terms \textit{online adjustment} and \textit{online adaptation} are interchangeable. They refer to the process of \textit{readjusting} the automation in response to the human adaptation in shared control interaction.} is treated in the state of the art.  

Therefore, this paper introduces an iterative model-based design approach that integrates online identification and controller design methods for shared control configurations. It offers a real-time implementation, combining and extending the approaches presented in \cite{2019_SolutionSetsInverse_inga} and \cite{2014_NecessarySufficientConditions_flad}. Firstly, this iterative method enables a more accurate controller design in case of a poorly pre-identified human operator. Secondly, it also facilitates an adaption of the global objective function during the operation of the shared control system. The highlights of this paper are:
\begin{itemize}
	\item Unifying the online identification \cite{2019_SolutionSetsInverse_inga} and design processes \cite{2014_NecessarySufficientConditions_flad} for a shared control configuration,
	\item Implementing the algorithm in real-time for practical application, and
	\item Conducting analysis using simulations and carrying out an experimental human-in-the-loop validation.
\end{itemize}
This paper has the following structure: In Section \ref{sec:Grundlagen}, the formulation of a shared control configuration as an LQ-differential game is introduced. 
Section \ref{sec:Method} presents the online iterative method of an LQ shared control, which includes the identification of the shared control configuration and the design of the automation. 
Section \ref{sec:Simulation} includes a simulation example for further analysis of the proposed concept, for which the ground truth of the human is available. The experimental setup and validation are given in Section \ref{sec:Experiment}. Furthermore, the limitations of the method are discussed. Finally, Section \ref{sec:Summary} summarizes the paper and provides an outlook on further research work.

\vspace*{-8mm}
\section{Problem formulation with LQ-Differential Games} \label{sec:Grundlagen}
It is widely assumed in the modeling of human actions that a cost function describes a their preferences and can be used to compute their actions. Consequently, literature often utilizes an optimal controller as a human model, as seen in various works such as \cite{2002_OptimalFeedbackControl_todorov, 2004_OptimalityPrinciplesSensorimotor_todorov, 2020_OptimalityPrinciplesHuman_wochner}. This optimality is utilized in this paper, to model the shared control interaction as an LQ differential game.

Consider a differential game with players indexed by $i = \{\mathrm{a},\mathrm{h}\} \in \pazocal{P}$, with the dynamic system $\sv{f} \in \mathbb{R}^n$ and the players' cost functions $J\play{i} \in \mathbb{R}$. For this work, the following assumptions are made:
\begin{itemize}
	\item The controlled system can be modeled as a linear system such that
	\begin{align} \label{eq:lin_system}
		\dot{\sv{x}}(t) &= \mathbf{A} \sv{x}(t) + \mathbf{B}\play{\mathrm{a}}\sv{u}\play{\mathrm{a}}(t) + \mathbf{B}\play{\mathrm{h}}\sv{u}\play{\mathrm{h}}(t)\\ \nonumber
		\sv{x}(0) &= \sv{x}_0,
	\end{align}
	where $\mathbf{A}$ and $\mathbf{B}\play{i}$ are the system matrix and the input matrices of the player $t$, respectively.
	The system is defined over the time horizon $t \in [0,\tau]$.
	\item The preferences of players are modeled as a quadratic cost function 
	\begin{align} \label{eq:quad_cost_f}
		J\play{i} &= \frac{1}{2}\int_{0}^{\tau} \sv{x}^\mathsf{T}(t) \mathbf{Q}\play{i} \sv{x}(t) \\ \nonumber
		&+ \sum_{i \in \pazocal{P}}{\sv{u}\play{j}}(t)^\mathsf{T}\mathbf{R}\play{ij} \sv{u}\play{j}(t) \text{ d}t,
	\end{align}
	where $i\in \pazocal{P}$. The matrices $\mathbf{Q}\play{i}$ and $\mathbf{R}\play{ij}$ are positive semi-definite and positive definite, respectively. The assumptions of a linear system (\ref{eq:lin_system}) and a quadratic cost function (\ref{eq:quad_cost_f}) are sufficient enough for many engineering applications see. e.g.~\cite{2005_LQDynamicOptimization_engwerda}.
	
\end{itemize}
These assumptions are also utilized in other state-of-the-art works, see e.g.~\cite{2005_LQDynamicOptimization_engwerda}. The Nash-equilibrium of this differential game is computed with the coupled optimization problem
\begin{equation}
	{\sv{u}\play{i}}^* = \underset{\sv{u}\play{i}}{\text{arg min }} J\play{i}(\cdot,\sv{u}\play{\neg i}),
\end{equation}
which obtains a feedback gain by LQ-games, where 
$i=\mathrm{a} \rightarrow \neg i=\mathrm{h}$ and vice versa. The inputs of the players are modeled with an optimal feedback gain
\begin{equation} \label{eq:control _law}
	\sv{u}\play{i} = - \sv{K}\play{i}\sv{x}.
\end{equation}
{The Nash-equilibrium of LQ-games is obtained from the solution of the coupled Riccati equation}
\begin{align} \nonumber \label{eq:ric_coup} 
	\vek{0} &=  \biggl(\vek{A}^\mathsf{T} \vek{P}\play{i} +  \vek{P}\play{i} \vek{A} + \vek{Q}\play{i} -
	\sum_{j \in \pazocal{P}} \vek{P}\play{i}\vek{S}\play{j}\vek{P}\play{j} \\ 
	&-\sum_{j \in \pazocal{P}} \vek{P}\play{j}\vek{S}\play{j}\vek{P}\play{i} + 
	\sum_{j \in \pazocal{P}} \vek{P}\play{j}\vek{S}\play{ij}\vek{P}\play{j} \biggr) \sv{x}(t), \; \forall i \in \pazocal{P}, \\ \nonumber
	\mathrm{where} \; \;&\vek{S}\play{j}  = \vek{B}\play{j} {\vek{R}\play{jj}}^{-1} {\vek{B}\play{j}}^\mathsf{T}	\; j \in \pazocal{P}, \\ \nonumber
	&\vek{S}\play{ij} = \vek{B}\play{j} {\vek{R}\play{jj}}^{-1} \vek{R}\play{ij} {\vek{R}\play{jj}}^{-1} {\vek{B}\play{j}}^\mathsf{T}\; j \in \pazocal{P}, i \neq j.
\end{align}
The control law of the player $i$ is calculated with the solution of \ref{eq:ric_coup}, $\mathbf{P}\play{i}$:
\begin{equation} \label{eq:control_law_players_nash}
	\sv{K}\play{i} = {\mathbf{R}\play{ii}}^{-1}{\mathbf{B}\play{i}}^\mathsf{T}\mathbf{P}\play{i},
\end{equation}
which leads the system in a Nash-equilibrium. In~\cite{2005_LQDynamicOptimization_engwerda}, Theorem 8.5. states that a feedback Nash-equilibrium is completely characterized by (\ref{eq:control_law_players_nash}), thus $\sv{K}\play{i}$ is sufficient for further system analysis. 

The feedback control law~$\sv{K}\play{i}$ can be computed if all the players' cost functions of the differential games are known. In the case of shared control interactions, the following challenges arise:
\begin{itemize}
	\item The cost function of the automation $J\play{\mathrm{a}}$ is to be designed by engineers, which depends on cost function of the human $J\play{\mathrm{h}}$.
	\item $J\play{\mathrm{h}}$ needs to be identified, which however changes with $J\play{\mathrm{a}}$.
\end{itemize}
Thus, $J\play{\mathrm{a}}$ and $J\play{\mathrm{h}}$ have a two-way effect, which leads to the a well-known \textit{chicken or the egg} causality dilemma. This causality dilemma of shared control system is not solved in the literature and not addressed by the existing research works. For this challenge, a novel solution is proposed in the next section.
\vspace*{-8mm}
\section{Novel Iterative Design of a Shared Control with Online Identification}\label{sec:Method}
\begin{figure}[t!]
	\centering
	\includegraphics[width=0.84\columnwidth]{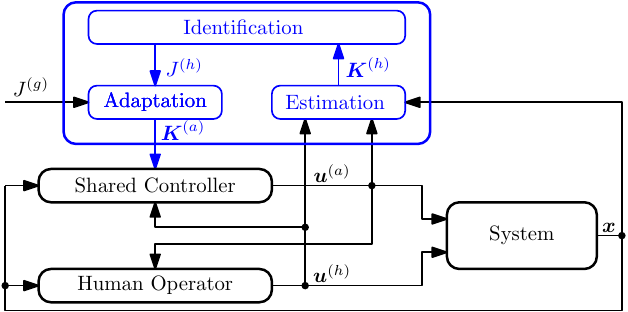}
	\caption{Block diagram of the adaptive shared control design with the online estimation}
	\label{fig:block_diag}
\end{figure}

The problem of identifying the human operator and designing the shared control\footnote{For the clarity note that in the following, the terms "automation" and "shared control" are used interchangeably.} is solved using an iterative online algorithm. This method is capable of handling changes in human preferences, such as alterations to the cost function, and also allows for online modification of the global objective function. 
\newpage
The steps of the adaptive design of the shared control are depicted in Figure~\ref{fig:block_diag}:
\begin{itemize}
	\item[1] Estimation of the control laws
	\item[2] Identification of the cost functions
	\item[3] Adaptation of the automation 
\end{itemize}
In the following, these steps are presented in detail.
\vspace*{-3mm}
\subsection{Estimation of the control law}
First, the feedback control law of the players is estimated from the measurement data. The identification of the control law for one player case can be found e.g.~in \cite{2008_AdaptiveControl_astrom} or \cite{2015_SolutionsInverseLQR_priess}. A least-square estimation is formulated, which yields an estimation of the control feedback gain of the player $i$, such that
\begin{align} \label{eq:K_schaetz}
	\hat{\sv{K}}\play{i} = \underset{\sv{K}\play{i}}{\text{arg min}}  \sum_{l = 1}^{M_k}  \left|\left|  \sv{u}\play{i}[l] - \hat{\sv{K}}\play{i}[l] \sv{x}[l] \right|\right|^2_2
\end{align}
where $[l]$ is the $l^\mathrm{th}$ measurement and $M_k$ the number of the measurements. The minimization problem (\ref{eq:K_schaetz}) is solved recursively with a recursive least-square (RLS) with an exponential forgetting factor \cite[Chapter 21]{2009_DigitalSignalProcessing_madisetti}. Using (\ref{eq:control _law}), the control law of the players is divided in
\begin{equation}
	\begin{bmatrix}
		u\play{i}_1\\
		\vdots \\
		u\play{i}_j \\
		\vdots \\
		u\play{i}_p
	\end{bmatrix}
	=
\underbrace{	\begin{bmatrix}
		{\sv{k}\play{i}_1}^\mathsf{T}   \\
		\vdots \\
		{\sv{k}\play{i}_j}^\mathsf{T}  \\
		\vdots \\
		{\sv{k}\play{i}_p}^\mathsf{T}
	\end{bmatrix}}_{\sv{K}\play{i}}
	\sv{x}
\end{equation}

The RLS provides for each measurement step the estimated feedback gain
\begin{align} \label{eq:comp_k_feed}
	\hat{\sv{k}}\play{i}_j [l+1] = \hat{\sv{k}}\play{i}_j [l] + \mathbf{W} \left( u_j\play{i}[l+1] - \sv{x}^\mathsf{T}[l+1] \hat{\sv{k}}\play{i}_j[l]\right),
\end{align}
where the weighting matrix $\mathbf{W}[l+1]$ is computed
\begin{align}
	\mathbf{W}[l+1] =& \frac{\mathbf{P}_{RLS}[l] \sv{x}[l+1]}{\lambda_\mathrm{f} + \sv{x}^\mathsf{T}[l]\mathbf{P}_{RLS}[l]\sv{x}[l+1]} \\
	\mathbf{P}_{RLS}[l+1] =& \frac{1}{\lambda_\mathrm{f}} \bigg( \mathbf{P}_{RLS}[l] - \mathbf{W}[l+1] \sv{x}^\mathsf{T}[l+1]\mathbf{P}_{RLS}[l] \bigg).
\end{align}
The parameter $\lambda_\mathrm{f}$ is the so-called forgetting factor, which governs the dynamics of the feedback gain $\sv{K}\play{i}$. Its choice is crucial since too large values and changes of the feedback gains are not tracked fast enough. On the other hand, too small values can prevent the convergence of the RLS.

Using \eqref{eq:comp_k_feed}, the feedback gains of the automation and the human in the shared control setup are estimated, which can be used for the online identification of their cost functions. The core idea of this identification method is presented in the next section.
\subsection{Identification of the cost functions of the players}

The second step of the proposed iterative design is the identification of the players' cost functions from their control laws. 
In this work, the identification method presented in \cite{2019_SolutionSetsInverse_inga} is used. 
The method is based on a reformulation of the algebraic Riccati equations into a system of equations of the form 
\begin{equation} \label{eq:inga_m1}
	\sv{M}\play{i}\sv{\theta}\play{i} = 0,
\end{equation}
where, the unknown parameters of the weighting matrices are summarized $\sv{\theta}\play{i}$\footnote{In case a shared control setup, $\sv{\theta}\play{i} = \left[\vek{Q}\play{i}, \; \vek{R}\play{i}\right]$ hold, where $i = \{\mathrm{a},\mathrm{h}\} \in \pazocal{P}$.}. The main benefit of this algorithm that $\sv{M}\play{i}$ depends only on the system matrices assumed to be known, as well as the estimated state feedbacks of the players $\hat{\sv{K}}\play{i}$. Such a split the know and unknown parts of the system enables a efficient identification of players' cost functions. In the following, the key aspects of the algorithm. For a deeper insight, it is referred to \cite{2019_SolutionSetsInverse_inga}.

To identify the parameters of the cost function, a residue is computed, such as ${r\play{i} = \sv{M}\play{i}\sv{\theta}\play{i}}$ using (\ref{eq:inga_m1}).
This residue serves as a measure of the deviation from the Nash equilibrium conditions.
Minimizing this residue 
\begin{align} \label{eq:inga_optim}
	\underset{\sv{\theta}\play{i}}{\mathrm{min }} &\;\;{\sv{\theta}\play{i}}^\mathsf{T}{\sv{M}\play{i}}^\mathsf{T} \sv{M}\play{i}\sv{\theta}\play{i}  \\ \nonumber
	\mathrm{ s.t. }\;\;& \sv{I}_L \sv{\theta}\play{i} > 0 \\ \nonumber
	& \sv{R}\play{ii} > 0
\end{align}
results in determining the cost function parameters for humans in the shared control configuration. 
The initial constraint in (\ref{eq:inga_optim}) is crucial for preventing trivial solutions such as $\sv{\theta}\play{i} = \sv{0}$. 
Note that the optimization problem (\ref{eq:inga_optim}) is convex due to ${\sv{M}\play{i}}^\mathsf{T}\sv{M}\play{i} \geq 0$.

In this paper, the real-time implementation of the identification method from \cite{2019_SolutionSetsInverse_inga} is provided in order to enable a online adaptation of the shared control.

\subsection{Adaptation of the Shared Control}
Following the core idea outlined in \cite{2014_SteeringDriverAssistance_flad}, a shared control design entails tailoring the overall behavior of a control loop to meet high-level requirements for a given human model. This adaptation is achieved through a optimization-based design of the shared control. The interaction between the shared control and the human is abstracted as a non-cooperative differential game\footnote{High-level requirements are stem from diverse sources, including upper management, customers, or end-users engaged in practical applications.}. Defining the most appropriate \textit{global objective function} $J\play{\mathrm{g}}$, which meets high-level requirements can be challenging the for real-world application. 

As proposed in \cite{2014_SteeringDriverAssistance_flad}, the global objective function $J\play{\mathrm{g}}$ is defined as quadratic function, such as
\begin{align} \label{eq:global_cost_quad}
	J\play{\mathrm{g}} = \frac{1}{2}\int_{0}^{\tau_\mathrm{end}}  \sv{x}(t)^\mathsf{T}  \vek{Q}\play{\mathrm{g}}\sv{x}(t)  +   \sum_{j \in \APlay} {\sv{u}\play{j}(t)}^\mathsf{T} \vek{R}\play{\mathrm{g}j}\sv{u}\play{j}(t)  \text{ d}t, 
\end{align}
where penalty matrices $\vek{Q}\play{g}$ and $\vek{R}\play{\mathrm{g}j}$ associated with system states and inputs, respectively, are derived from high-level requirements. The matrices in the objective function (\ref{eq:global_cost_quad}) are assumed to be diagonal, such as
\begin{align*}
	\vek{Q}\play{\mathrm{g}} &= \mathrm{diag}[q\play{\mathrm{g}}_1,q\play{\mathrm{g}}_2,...,q\play{\mathrm{g}}_n], \\ 
	\vek{R}\play{\mathrm{g}}&=\mathrm{diag}[r\play{\mathrm{g}}_1,r\play{\mathrm{g}}_2,...,r\play{\mathrm{g}}_{Np}].
\end{align*}
This diagonalization, a common practice in optimal control theory, simplifies the interpretation of the cost function by eliminating mixed terms outside the diagonal, as discussed in \cite{bryson2018applied}.

To obtain the optimal parameters of the shared control 
$\sv{\theta}\play{\mathrm{a}} = \left[\vek{Q}\play{\mathrm{a}}, \; \vek{R}\play{\mathrm{a}}\right]$ 
and to compute its feedback gain $\sv{K}\play{\mathrm{a}}$, $J\play{\mathrm{g}}$ is optimized such as
\begin{subequations} \label{eq:flad_global_lq}
	\begin{align} 
		{\sv{\theta}\play{\mathrm{a}}}^* =&  \,\underset{\sv{\theta}\play{\mathrm{a}}}{\mathrm{arg \, min }}  \; J\play{g} \left(t,\tau_\mathrm{end}, \sv{x}(t),{\sv{u}\play{\mathrm{h}}}^*(t), {\sv{u}\play{\mathrm{a}}}^*(t) , \sv{\theta}\play{\mathrm{a}} \right), \\ \nonumber
		& \;  \mathrm{w.r.t} \; \forall i \in \left\{\mathrm{h},\mathrm{a}\right\} \\
		\vek{0} =& \vek{A}^\mathsf{T} \vek{P}\play{i} +  \vek{P}\play{i} \vek{A} + \vek{Q}\play{i} -
		\sum_{j \in \pazocal{P}} \vek{P}\play{i}\vek{S}\play{j}\vek{P}\play{j} \\ \nonumber 
		-&\sum_{j \in \pazocal{P}}\vek{P}\play{j}\vek{S}\play{j}\vek{P}\play{i} + 
		\sum_{j \in \pazocal{P}}\vek{P}\play{j}\vek{S}\play{ij}\vek{P}\play{j},
	\end{align}
\end{subequations}
which yields the desired optimal parameters with respect to the identify human. 
However, in the event an alteration in human behavior, specifically changes in their cost function, it becomes necessary to adapt the optimal parameters of the shared control accordingly.

This research has led to the development of a real-time implementation for \ert{eq:flad_global_lq}, allowing for the dynamic online design of model-based shared control.

\subsection{Analysis for Practical Usage} \label{sec:practical_anal}
In order to enable a practical use of the proposed online shared control design, the following criteria need to be taken into account: The online shared control 
\begin{itemize}
	\item[a)] needs to lead to a improvement of the overall performance,
	\item[b)] must run in real time and
\end{itemize}
Since, the adaptation loop (see blue part in Figure~\ref{fig:block_diag}) is an additional feedback in the system, a convergence analysis of the adaptive shared control is essential.

\paragraph{Convergence of the feedback gain estimation \ert{eq:comp_k_feed}}
The control law of the human~$\hat{\sv{K}}\play{h}$ is obtained from the RLS-estimation. The ground truth of the optimal feedback gain ${\sv{K}\play{h}}^*$ is computed from the optimization \ert{eq:flad_global_lq}. Thus, the error
\begin{equation} \label{eq:human_controller_errror}
	\sv{e}\play{h}_{\sv{K}}(t) = \left|\left|{\sv{K}\play{h}}^*(t) - \hat{\sv{K}}\play{h}(t)\right|\right|_2
\end{equation}
can characterize the convergence of the feedback gain estimation. If the error $\sv{e}_c\play{h}(t)$ convergences to zero, the identification of the human's cost function and the update of the automation provides a good estimation of the human control law indicating that the parameters of the RLS-estimation are chosen properly. 

\paragraph{Convergence of the identification \ert{eq:inga_optim}}
The convergence of (\ref{eq:inga_optim}) is ensured through the quadratic structure as stated in \cite{2019_SolutionSetsInverse_inga}, since ${\sv{M}\play{i}}^\mathsf{T}\sv{M}\play{i} > 0$ holds and therefore, the obtained solution is unique. However, in practical application, ${\sv{M}\play{i}}^\mathsf{T}\sv{M}\play{i} \geq 0$ holds, meaning that the initial values of optimization \ert{eq:inga_optim} has in impact on the identification. As $\vek{M}\play{i}$ is influenced by the system dynamics, a separate examination is required for each system.

\paragraph{Convergence of the adaptation \ert{eq:flad_global_lq}}
In the following, the control law of the automation is proofed weather the updated control law leads to a stable closed loop system dynamics. The closed loop-system dynamics is
\begin{equation}
	\mathbf{A}_\mathrm{cl}(t) = \left(\mathbf{A} - \mathbf{B}\play{\mathrm{a}} \sv{K}\play{\mathrm{a}}_\mathrm{adap}(t) - \mathbf{B}\play{\mathrm{h}} \hat{\sv{K}}\play{\mathrm{h}}(t)\right),
\end{equation}
where the eigenvalues characterize the stability of the system.	For each time steps, the eigenvalues $\Lambda\left(\vek{A}_\mathrm{cl}(t)\right)$ need to be computed to verify convergence of the adaptation.

\section{Simulation Analysis} \label{sec:Simulation}
In the forthcoming section, the efficiency of the proposed shared control and further analysis are conducted through simulations.
Simulations offer a distinct advantage as the ground truth of human behavior can be pre-defined, enabling an in-depth examination of the identification results given by \ert{eq:inga_optim}. The simulations are carried out using Matlab/Simulink, R2021b \cite{matlab_2021}.

The numerical example is drawn from the practical application of vehicle manipulators utilized in road maintenance tasks. Such a scenario involves a shared control, where both a human and automation jointly operate two subsystems~-~the vehicle and the manipulator. Additional details can be found in \cite{2019_ModelPredictiveControl_varga} or \cite{2023_LimitedInformationShared_vargaa_Diss}. It is crucial to note that the stability of the overall system relies on the shared efforts of both the human (controlling the manipulator) and the automation (steering the vehicle). The absence of either component would result in system instability. In contrast to \cite{2023_LimitedInformationShared_vargaa_Diss}, the proposed approach assumes the measurement and usage of all system states for shared control adaptation.

\subsection{Differential Game Formulation}
For this technical application of a vehicle manipulator, linear system dynamics are assumed, such as \ert{eq:lin_system}, where the system and input matrices are
\begin{equation} \label{eq:ABs4_lin_sys}
\vek{A} = \begin{bmatrix}
	-0.1 & 0 & 0 \\
	0 & 0 & 0.9 \\
	0 & 0 & 0
\end{bmatrix}, \; \; \vek{B}\play{\mathrm{h}} = \begin{bmatrix}
		0.85 \\
		0 \\
		0
	\end{bmatrix} \; \mathrm{and} \; \;
	\vek{B}\play{\mathrm{a}} = \begin{bmatrix}
		1.95 \\
		0 \\
		1.25
	\end{bmatrix}.
\end{equation}
The cost function of the human is in a quadratic form \ert{eq:quad_cost_f}, where the weighting matrices are
\begin{equation} \label{eq:J_hum_Num}
	\vek{Q}\play{\mathrm{h}} = \mathrm{diag}\left(50,\; 0.2,\; 0.2\right) \;  \mathrm{and} \; \vek{R}\play{\mathrm{h}} = 1
\end{equation}
and the global objective function is
\begin{equation} \label{eq:J_glob_Num}
	\vek{Q}\play{\mathrm{g}} = \mathrm{diag}\left(35,\;1,\;3\right) \;  \mathrm{and} \; \vek{R}\play{\mathrm{h}} = \mathrm{diag}\left(1,\;1\right).
\end{equation}

The linear system dynamics are defined in an error-frame relative to both the references of the vehicle and the manipulator.
The human operator strives for precise tracking of the manipulator reference.
The automation's primary goal is to assist the human, indicated by the parameters of $\vek{Q}\play{\mathrm{g}}$ in \ert{eq:J_glob_Num}.
It's important to note that precise tracking of the vehicle reference is not crucial, resulting in considerably smaller weights in $J\play{\mathrm{g}}$.
 
In order to obtain the parameters of the automation, the optimization \ert{eq:flad_global_lq} is carried out, offline using the parameters from \ert{eq:J_hum_Num} and \ert{eq:J_glob_Num}, which yields
\begin{subequations}
\begin{align} \label{eq:K_offline}
 \sv{K}\play{h} = &\left[     3.16, \;  -0.69, \;  -1.88\right]\\
  \sv{K}\play{a} =&\left[     4.39, \;  0.69, \;  1.62\right]
\end{align}
\end{subequations}
for the human and the automation. 
In order to model the changes of the human, the cost function is changed in the middle of the simulation to 
\begin{equation} \label{eq:q_hum_ch}
	\vek{Q}\play{\mathrm{h}}_\mathrm{ch} =\left(0.5,\;0.2,\;0.2\right).
\end{equation}
Consequently, the feedback gains of the human change to
\begin{equation}
	\sv{K}\play{\mathrm{h}}_\mathrm{ch} = \left[0.72,\;   -0.38,\;   -1.13\right]
\end{equation}
where the "ch" index denotes \textit{modified} values
Such alterations in human awareness may stem from factors such as fatigue or variations in environmental conditions like dusk or fog.
The forgetting factor is chosen to $\lambda_\mathrm{f} = 0.985$ aiming for an optimal trade-off between considering changes and ensuring convergence stability.

For the objective comparison between the adaptive and non-adaptive shared control, the root-mean-square error (RMSE) of the manipulator from its reference 
\begin{equation}
	|d_\mathrm{m}| = \sqrt{\frac{1}{M}\sum_{k}^{M} d^2_\mathrm{m}[k]}.
\end{equation}
is chosen as a measure.

\subsection{Simulation Results and Discussion}
In the simulation scenario, the same track is driven twice by the vehicle manipulator, in which the cost function of the human operator changes from \ert{eq:J_hum_Num} to \ert{eq:q_hum_ch}.
\begin{figure}[!b]
	\centering
	\includegraphics[width=1.0\linewidth,height=\figHight cm]{traj_comp}
	\caption{Comparison of the trajectories obtained from the non-adaptive shared control (solid lines) and from the proposed adaptive case (dashed lines)}
	\label{fig:comp_trajs}
\end{figure}
The simulation scenario takes $120\,$s long and the cost function change happens $t=60\,$s. 

In the first case, the automation is not adapted, the same feedback gain is used for the whole simulation, which is computed offline in advance.
In the second case, in every $1\,$s, the automation is adapted according to \ert{eq:flad_global_lq}.

The resulting trajectories of the vehicle manipulator are given in Figure \ref{fig:comp_trajs}, showcasing the performance with and without adaptation following the transition from $\vek{Q}\play{\mathrm{h}}$ to $\vek{Q}\play{\mathrm{h}}_\mathrm{ch}$.
It can be seen that by using the proposed adaptive shared control, more accurate tracking with the manipulator is possible. 
This scenario models the practical case, in which the human operator cannot concentrate on their task, therefore more support from the automation is required. 
In this case, using the adaptation has benefits and increases the overall performance of the shared control system.

\begin{figure}[!t]%
	\centering
	\includegraphics[width=1.0\linewidth,height=\figHight cm]{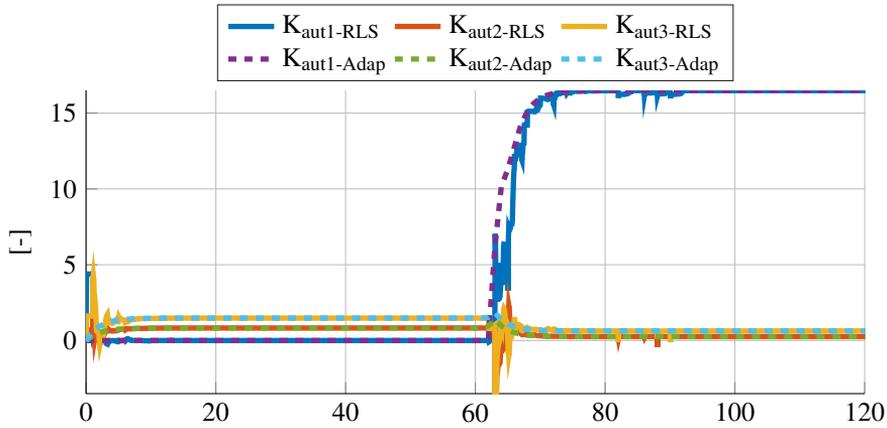}
	\caption{The feedback gains of the human operator}
	\label{fig:hum_K_changes}
\end{figure}

\begin{figure}[!t]
	\centering
	\includegraphics[width=1.0\linewidth,height=\figHight cm]{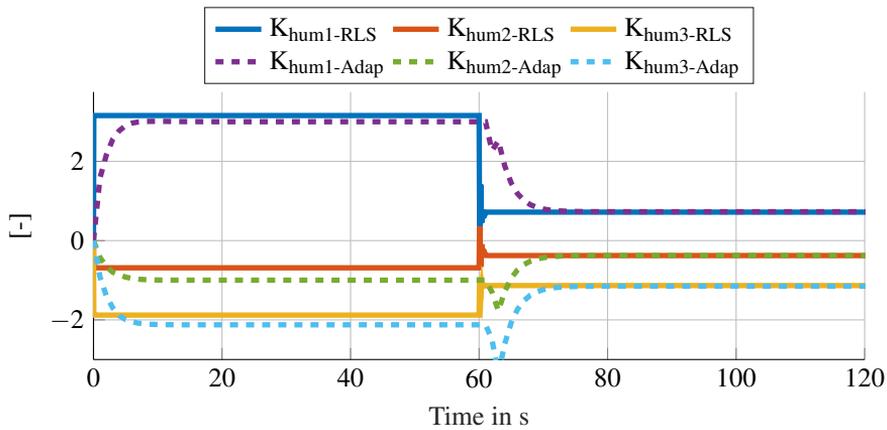}
	\caption{The feedback gains of the shared control}
	\label{fig:aut_K_changes}
\end{figure}

The feedback gains of both the human and shared control are illustrated in Figures \ref{fig:hum_K_changes} and \ref{fig:aut_K_changes}, respectively. These gains are either identified using \ert{eq:K_schaetz} or calculated from \ert{eq:flad_global_lq}, \ert{eq:control_law_players_nash}. 
The decreasing differences $\sv{e}\play{h}_{\sv{K}}(t)$ over time indicate that the identification of the feedback gain converges in accordance with \ert{eq:human_controller_errror}. 
Furthermore, as the human's feedback gain decreases at $t= 60\,$s, the automation's feedback increases, providing more support for the human operator in their task.

\begin{figure}[!t]%
	\centering
	\includegraphics{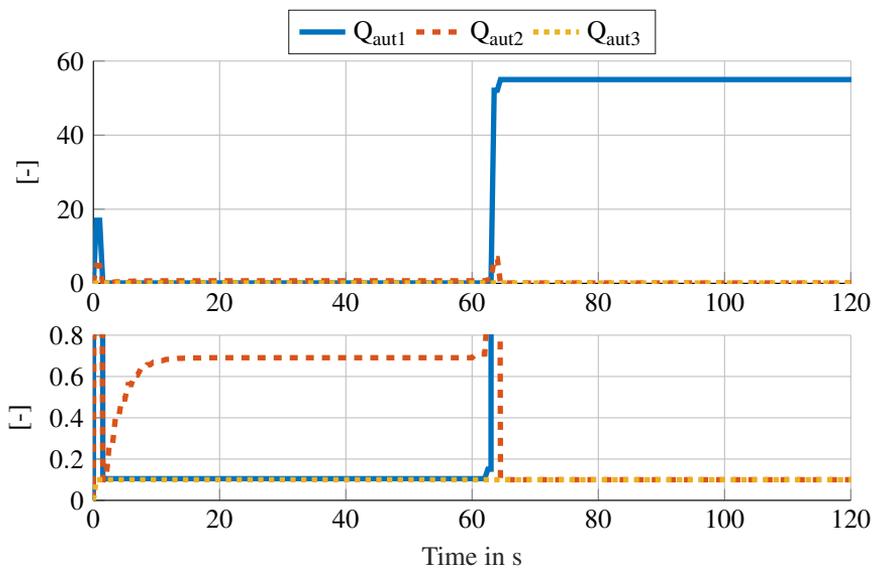}
	\caption{Changes of the diagonal elements of the automation cost functions\\ The diagram at the bottom provides a more detailed view, focusing on smaller values}
	\label{fig:Q_aut_sim}
\end{figure}

In Figure~\ref{fig:Q_aut_sim}, the adapted cost function components of the automation are given, from which it can be seen that $Q_1$ is increased meaning that stronger support from the automation is necessary in order to obtain an optimal $J\play{\mathrm{g}}$ from \ert{eq:flad_global_lq}

\begin{figure}[!t]
	\centering
	\includegraphics[width=1.0\linewidth,height=\figHight cm]{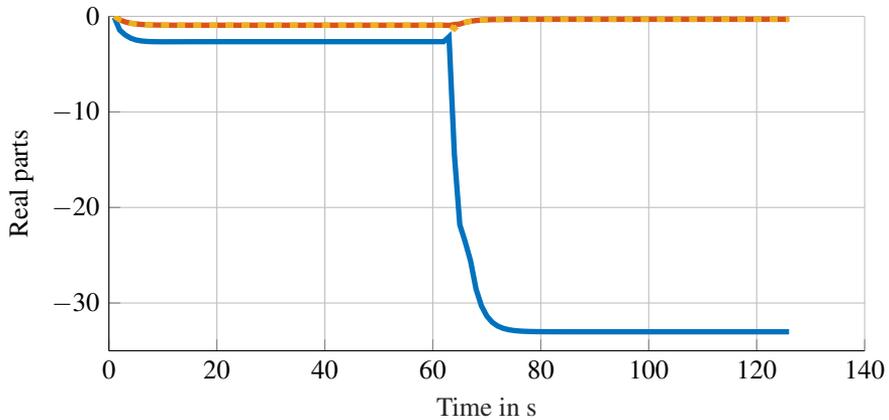}
	\caption{The eigenvalues of the overall system}
	\label{fig:eigen_v}
\end{figure}

In order to analyze the stability of the overall system, the real parts of its eigenvalues are given in Figure~\ref{fig:eigen_v}, which shows that the system remains stable even after the adaptation of the automation: The largest values are 
$$
\mathrm{max}\Bigl( \Lambda\left(\vek{A}_\mathrm{cl}\right)\Bigl) = \left[-1.44, -0.31, -0.31\right].
$$
Thus, the adaptation \ert{eq:flad_global_lq} leads to a stable overall system behavior.
The measures after the adaptation yielded
$|d_\mathrm{m}|_{w/ \, adapt} =1.87 \cdot 10^{-5}$ and $|d_\mathrm{m}|_{w/o \, adapt} =2.03 \cdot 10^{-4}$, showing that the adaptation leads to a more accurate tracking compared to the non-adaptive case.
To summarize the simulation analysis, the results provide the first proof of the concept of an adaptive shared control, which leads to a control behavior adapted to the human operator.
\vspace*{-8mm}
\section{Experimental Validation} \label{sec:Experiment}
In order to demonstrate the functionality of the proposed online adaptive shared control in a real-time application, the system is validated in an experimental setup.
\subsection{Technical System}
\begin{figure}[!t]
	\centering
	\includegraphics[width=0.95\linewidth]{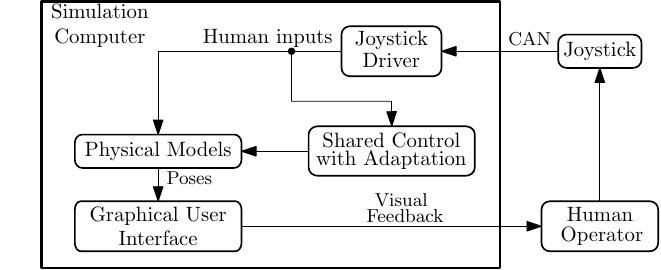}
	\caption{Software and hardware structure of the simulator for the experimental validation of the adaptive shared control concept.}
	\label{fig:soft_hard_struct}
\end{figure}

In this experimental setup, a human controls the manipulator with a joystick based on the visual feedback from a simplified graphical user interface of the scenario. The system setup is depicted in Figure~\ref{fig:soft_hard_struct}, which is implemented using the common Robot Operating System (ROS).

The components given in Figure \ref{fig:soft_hard_struct} are implemented as ROS nodes. 
The adaptive shared control has been auto-generated from Matlab as a C++ ROS node. The generated code is run at 25 Hz in real time.
The scenario includes different types of trajectories with sudden changes and smooth paths.

\subsection{Measurement Results}
In Figure \ref{fig:Exp_res_traj}, a section of the resulting trajectories is given, in which the adaptive (dashed lines) and the non-adaptive (solid lines) shared controls are compared. It can be seen that the human operator can track the reference of the manipulator more accurately than with the non-adaptive controller since the deviations from the manipulator's reference are smaller. It needs to be mentioned that the tracking of the vehicle reference had a low priority meaning that the tracking of the reference of the vehicle is not evaluated.
\begin{figure}[!t]
	\centering
	\includegraphics[width=1.0\linewidth,height=7.5cm]{exp_traj}
	\caption{Resulting trajectories from the experiment, where the adaptive shared control (dashed lines) is compared with the non-adaptive case. References are transparent solid lines.}
	\label{fig:Exp_res_traj}
\end{figure}
\vspace*{-3mm}
\subsection{Discussion on the Optimality Principle of the Human}
The simulations and the experimental validation serve as the initial proofs of concept for the proposed adaptive shared control, demonstrating its practicality in human-in-the-loop scenarios.
The concept operates based on the optimality principle of the human operator, implying that the feedback gains are valid when the human behaves optimally. Thus, for proper functionality, a reasonable human behavior is required. Furthermore, the validity of the cost function's structure needs to be examined in accordance to~\cite{karg2024trustworthiness}.

Another limitation of the proposed concept is the lack of consideration for the variability of the human, which could be a crucial aspect for acceptance, as indicated in current research, e.g.~\cite{2023_BiLevelBasedInverseStochastic_karg}.
In the case of an inadequate estimation of the human's feedback gain $\sv{K}\play{\mathrm{h}}$, it is essential to maintain the stability of the human-in-the-loop system. Conducting stability analysis in such scenarios requires additional methods from the literature, as discussed in~\cite{2017_StabilityAnalysisHumanintheLoop_yousefi}.
Notwithstanding these limitations, this experimental validation establishes a solid foundation, affirming that the proposed adaptive shared control is applicable in human-in-the-loop systems.

\section{Conclusion and Outlook} \label{sec:Summary}
This paper presents the first proof of concept of an adaptive shared control based on LQ-differential games. 
The concept unifies the identification and design processes for a shared control configuration based on earlier works. 
Furthermore, the algorithm is implemented in real-time for practical application.
Analyses using simulations are conducted, in which the advantages of the proposed adaptive shared control are highlighted.
Finally, the practical usability of the proposed concept is validated with an experimental human-in-the-loop setup.

In future work, the choice of $J\play{\mathrm{g}}$ and its impact on the computational performance and on the overall results will be investigated in detail. Furthermore, a study with numerous test subjects is planned, in order to show a statistically relevant benefit of the proposed adaptive shared control.

{{
	
\bibliographystyle{acta}
%\bibliography{Literatur_SACI}

\begin{thebibliography}{10}

\bibitem{2016_SharedControlSharp_flemisch}
F.~Flemisch, D.~Abbink, M.~Itoh, M.-P. {Pacaux-Lemoine}, and G.~We{\ss}el.
\newblock Shared control is the sharp end of cooperation: {{Towards}} a common
  framework of joint action, shared control and human machine cooperation.
\newblock {\em IFAC-PapersOnLine}, 49(19):72--77, 2016.

\bibitem{1983_IroniesAutomation_bainbridge}
L.~Bainbridge.
\newblock Ironies of automation.
\newblock {\em Automatica}, 19(6):775--779, November 1983.

\bibitem{1980_FlightdeckAutomationPromises_wiener}
E.~L. Wiener and R.~E. Curry.
\newblock Flight-deck automation: Promises and problems.
\newblock {\em Ergonomics}, 23(10):995--1011, November 1980.

\bibitem{9369899}
A.~M.~R. Lazcano, T.~Niu, X.~C. Akutain, D.~Cole, and B.~Shyrokau.
\newblock Mpc-based haptic shared steering system: A driver modeling approach
  for symbiotic driving.
\newblock {\em IEEE/ASME Transactions on Mechatronics}, 26(3):1201--1211, 2021.

\bibitem{2023_Na_Cole}
X.~Na and D.~J. Cole.
\newblock Experimental evaluation of a game-theoretic human driver steering
  control model.
\newblock {\em IEEE Transactions on Cybernetics}, 53(8):4791--4804, 2023.

\bibitem{2022_SharedControlRobot_gottardi}
A.~Gottardi, S.~Tortora, E.~Tosello, and E.~Menegatti.
\newblock Shared {{Control}} in {{Robot Teleoperation With Improved Potential
  Fields}}.
\newblock {\em IEEE Trans. Human-Mach. Syst.}, 52(3):410--422, June 2022.

\bibitem{2022_IntelligentInteractionFramework_du}
G.~Du, Y.~Deng, W.~W.~Y. Ng, and D.~Li.
\newblock An {{Intelligent Interaction Framework}} for {{Teleoperation Based}}
  on {{Human-Machine Cooperation}}.
\newblock {\em IEEE Trans. Human-Mach. Syst.}, 52(5):963--972, October 2022.

\bibitem{2022_ReviewHumanMachine_yanga}
C.~Yang, Y.~Zhu, and Y.~Chen.
\newblock A {{Review}} of {{Human}}{\textendash}{{Machine Cooperation}} in the
  {{Robotics Domain}}.
\newblock {\em IEEE Trans. Human-Mach. Syst.}, 52(1):12--25, February 2022.

\bibitem{2013_HapticSupportBimanual_kuiper}
R.~J. Kuiper, J.~C. Frumau, F.~C. van~der Helm, and D.~A. Abbink.
\newblock Haptic {{Support}} for {{Bi-manual Control}} of a {{Suspended Grab}}
  for {{Deep-Sea Excavation}}.
\newblock In {\em 2013 {{IEEE International Conference}} on {{Systems}},
  {{Man}}, and {{Cybernetics}}}, pages 1822--1827, {Manchester}, October 2013.
  {IEEE}.

\bibitem{2017_NeuromuscularSystemBasedTuningHaptic_smisek}
J.~Smisek, E.~Sunil, M.~M. Van~Paassen, D.~A. Abbink, and M.~Mulder.
\newblock Neuromuscular-{{System-Based Tuning}} of a {{Haptic Shared Control
  Interface}} for {{UAV Teleoperation}}.
\newblock {\em IEEE Trans. Human-Mach. Syst.}, 47(4):449--461, August 2017.

\bibitem{2023_LimitedInformationShared_varga}
B.~Varga, J.~Inga, and S.~Hohmann.
\newblock Limited {{Information Shared Control}}: {{A Potential Game
  Approach}}.
\newblock {\em IEEE Trans. Human-Mach. Syst.}, 53(2):282--292, April 2023.

\bibitem{2012_HapticSharedControl_abbink}
D.~A. Abbink, M.~Mulder, and E.~R. Boer.
\newblock Haptic shared control: Smoothly shifting control authority?
\newblock {\em Cognition, Technology \& Work}, 14(1):19--28, March 2012.

\bibitem{2002_OptimalFeedbackControl_todorov}
E.~Todorov and M.~I. Jordan.
\newblock Optimal feedback control as a theory of motor coordination.
\newblock {\em Nature Neuroscience}, 5(11):1226--1235, November 2002.

\bibitem{1969_HumanOptimalController_baron}
S.~Baron and D.~Kleinman.
\newblock The {{Human}} as an {{Optimal Controller}} and {{Information
  Processor}}.
\newblock {\em IEEE Transactions on Man Machine Systems}, 10(1):9--17, March
  1969.

\bibitem{2002_HumanMotionPlanning_loa}
J.~Lo, G.~Huang, and D.~Metaxas.
\newblock Human {{Motion Planning Based}} on {{Recursive Dynamics}} and
  {{Optimal Control Techniques}}.
\newblock {\em Multibody System Dynamics}, 8(4):433--458, 2002.

\bibitem{2010_HumanHumanoidLocomotion_mombaur}
K.~Mombaur, A.~Truong, and J.-P. Laumond.
\newblock From human to humanoid locomotion\textemdash an inverse optimal
  control approach.
\newblock {\em Autonomous Robots}, 28(3):369--383, April 2010.

\bibitem{2015_SolutionsInverseLQR_priess}
M.~C. Priess, R.~Conway, J.~Choi, J.~M. Popovich, and C.~Radcliffe.
\newblock Solutions to the {{Inverse LQR Problem With Application}} to
  {{Biological Systems Analysis}}.
\newblock {\em IEEE Transactions on Control Systems Technology}, (2), 2015.

\bibitem{2009_NashEquilibriaMultiAgent_braun}
D.~A. Braun, P.~A. Ortega, and D.~M. Wolpert.
\newblock Nash {{Equilibria}} in {{Multi-Agent Motor Interactions}}.
\newblock {\em PLoS Computational Biology}, 5(8):e1000468, August 2009.

\bibitem{2014_SteeringDriverAssistance_flad}
M.~Flad, J.~Otten, S.~Schwab, and S.~Hohmann.
\newblock Steering driver assistance system: {{A}} systematic cooperative
  shared control design approach.
\newblock In {\em 2014 {{IEEE International Conference}} on {{Systems}},
  {{Man}}, and {{Cybernetics}} ({{SMC}})}, pages 3585--3592, {San Diego, CA,
  USA}, October 2014. {IEEE}.

\bibitem{2019_SolutionSetsInverse_inga}
J.~Inga, E.~Bischoff, T.~L. Molloy, M.~Flad, and S.~Hohmann.
\newblock Solution {{Sets}} for {{Inverse Non-Cooperative Linear-Quadratic
  Differential Games}}.
\newblock {\em IEEE Control Systems Letters}, 3(4):871--876, October 2019.

\bibitem{2022_inga_application}
P.~Franceschi, N.~Pedrocchi, and M.~Beschi.
\newblock Inverse optimal control for the identification of human objective: a
  preparatory study for physical human-robot interaction.
\newblock In {\em 2022 IEEE 27th International Conference on Emerging
  Technologies and Factory Automation (ETFA)}, pages 1--6, 2022.

\bibitem{2023_IdentificationHumanControl_}
P.~Franceschi, N.~Pedrocchi, and M.~Beschi.
\newblock Identification of human control law during physical human–robot
  interaction.
\newblock {\em Mechatronics}, 92:102986, 2023.

\bibitem{Preferencemodelling_Palo}
P.~Franceschi, M.~Maccarini, D.~Piga, M.~Beschi, and L.~Roveda.
\newblock Human preferences' optimization in phri collaborative tasks.
\newblock In {\em 2023 20th International Conference on Ubiquitous Robots
  (UR)}, pages 693--699, 2023.

\bibitem{2014_NecessarySufficientConditions_flad}
M.~Flad, J.~Otten, S.~Schwab, and S.~Hohmann.
\newblock Necessary and sufficient conditions for the design of cooperative
  shared control.
\newblock In {\em 2014 {{IEEE International Conference}} on {{Systems}},
  {{Man}}, and {{Cybernetics}} ({{SMC}})}, pages 1253--1259, {San Diego, CA,
  USA}, October 2014. {IEEE}.

\bibitem{2004_OptimalityPrinciplesSensorimotor_todorov}
E.~Todorov.
\newblock Optimality principles in sensorimotor control.
\newblock {\em Nature Neuroscience}, 7(9):907--915, September 2004.

\bibitem{2020_OptimalityPrinciplesHuman_wochner}
I.~Wochner, D.~Driess, H.~Zimmermann, D.~F.~B. Haeufle, M.~Toussaint, and
  S.~Schmitt.
\newblock Optimality {{Principles}} in {{Human Point-to-Manifold Reaching
  Accounting}} for {{Muscle Dynamics}}.
\newblock {\em Frontiers in Computational Neuroscience}, 14:38, May 2020.

\bibitem{2005_LQDynamicOptimization_engwerda}
J.~Engwerda.
\newblock {\em {{LQ Dynamic Optimization}} and {{Differential Games}}}.
\newblock {The Netherlands}, tilburg {University}, edition, 2005.

\bibitem{2008_AdaptiveControl_astrom}
K.~J. {\AA}str{\"o}m and B.~Wittenmark.
\newblock {\em Adaptive Control}.
\newblock {Dover Publications}, {Mineola, N.Y.}, 2nd ed., dover ed edition,
  2008.

\bibitem{2009_DigitalSignalProcessing_madisetti}
V.~Madisetti.
\newblock {\em Digital {{Signal Processing Fundamentals}}}, volume 20094251 of
  {\em Electrical {{Engineering Handbook}}}.
\newblock {CRC Press}, November 2009.

\bibitem{bryson2018applied}
A.~E. Bryson and Y.-C. Ho.
\newblock {\em Applied Optimal Control: Optimization, Estimation, and Control}.
\newblock {Routledge}, 2018.

\bibitem{matlab_2021}
T.~M. Inc.
\newblock Matlab version: 9.11.0 (r2021b), 2021.

\bibitem{2019_ModelPredictiveControl_varga}
B.~Varga, S.~Meier, S.~Schwab, and S.~Hohmann.
\newblock Model {{Predictive Control}} and {{Trajectory Optimization}} of
  {{Large Vehicle-Manipulators}}.
\newblock In {\em 2019 {{IEEE International Conference}} on {{Mechatronics}}
  ({{ICM}})}, pages 60--66, {Ilmenau, Germany}, March 2019. {IEEE}.

\bibitem{2023_LimitedInformationShared_vargaa_Diss}
B.~Varga.
\newblock {\em Limited {{Information Shared Control}} and Its {{Applications}}
  to {{Large Vehicle Manipulators}}}.
\newblock {Karlsruher Institut f{\"u}r Technologie (KIT)}, 2023.

\bibitem{karg2024trustworthiness}
P.~Karg, A.~Kienzle, J.~Kaub, B.~Varga, and S.~Hohmann.
\newblock {Trustworthiness} of {Optimality Condition Violation} in {Inverse Optimal 
Control Methods Based} on the {Minimum Principle}, {\em preprint arXiv:2402.03157} 2024.

\bibitem{2023_BiLevelBasedInverseStochastic_karg}
P.~Karg, M.~Hess, B.~Varga, and S.~Hohmann.
\newblock  Bi-{{Level-Based Inverse Stochastic Optimal Control}}.
\newblock Accepted for publication in {\em IEEE European Control Conference 2024}. IEEE.


\bibitem{2017_StabilityAnalysisHumanintheLoop_yousefi}
E.~Yousefi, Y.~Yildiz, R.~Sipahi, and T.~Yucelen.
\newblock Stability {{Analysis}} of a {{Human-in-the-Loop Telerobotics System}}
  with {{Two Independent Time-Delays}}.
\newblock {\em IFAC-PapersOnLine}, 50(1):6519--6524, July 2017.

\end{thebibliography}

}
}
\end{document}